\documentclass[10pt,aps,pre,superscriptaddress,twocolumn]{revtex4-2}
\usepackage{graphicx}
\usepackage{dcolumn}
\usepackage{bm}
\usepackage{color}
\usepackage[utf8]{inputenc} 
\usepackage[english]{babel}
\usepackage{mathtools}
\usepackage{amsfonts}
\usepackage{amsmath}
\usepackage{amsthm}
\usepackage{color}
\usepackage{multirow}
\usepackage{rotating}
\begin{document}

\title{An entropic approach to analyze phase transitions in the $q=3$ Potts model}

\author{L. S. Ferreira}
\email{ferreira\_s\_lucas@ufg.br}
\affiliation{Departamento de Física, Universidade Federal de Minas Gerais, Av. Antônio Carlos 6627, 31270-901, Belo Horizonte, Minas Gerais, Brazil}%

\author{L. N. Jorge}
\affiliation{Instituto Federal do Mato Grosso - Campus C{\'a}ceres, Av. dos Ramires s/n, 78200-000, C{\'a}ceres, MT, Brazil}

\author{C. J. DaSilva}
\affiliation{Instituto Federal de Goi\'as, Rua 76, Centro, Goi\^ania - GO, Brazil}

\author{A. A. Caparica}
\affiliation{Instituto de F\'{i}sica, Universidade Federal de Goi\'{a}s, Av. Esperan\c{c}a s/n, 74.690-900, Goi\^{a}nia, GO, Brazil}%



\date{\today}

\begin{abstract}

Boltzmann's microcanonical entropy is the link between statistical physics and thermodynamics, forasmuch as the behavior of any thermodynamic quantity is directly related to the number of microscopic configurations. Accordingly, in this work, we investigate the behavior of the logarithm of the density of states of the three-state Potts model with an external field applied to one of the states using joint entropic simulations based on the Wang-Landau algorithm. Our analysis reveals that the microcanonical entropy curve is detachable, and each resulting path is related to the formation of clusters. Such a description is consistent with the energy-entropy argument related to the inception of a phase transition. When the external field is reversed and strong, the observed phase transition is from an ordered configuration to cluster formations. The behavior of the microcanonical inverse temperature indicates both first and second-order phase transitions occurring at different temperatures for high values of the external field.
\end{abstract}

\maketitle

\section{Introduction}


Phase transitions have been attracting attention for a long time. Its aspects are outward from microscopic to cosmic dimensions \cite{Linde1979,Witten1984,Mazumdar2019}. However, its description is not as prompt as its observation. For instance, the theoretical description of what happens at the triple point of water or its critical behavior is not straightforward\cite{Fisher1974}. The fluctuations are immense, and the visualization of a phase transition is subtle. 

To tackle phase transitions theoretically, one can define an order parameter to characterize a particular phase\cite{Landau1980}. This quantity has a non-zero value in an ordered phase and is null in the disordered one. However, for a second-order phase transition, the order parameter only signals the loss of the system order, unable to determine which configuration the system attains.


It is well-known that at the critical point the system comprises two indistinguishable phases, as a result of the formation of clusters of different sizes. In a real system, this discretion relies on the visualization of critical opalescence due to the scattering of light at all frequencies, due to the different sizes of the phases\cite{Gopal2000,Williamson2021}. Configurations at the critical point certainly do not satisfy the order parameter, so they can easily be confused with disordered configurations. A common feature at the critical point is the dependence of thermodynamic properties on the size of the system, called the finite-size effect\cite{Fisher1972,Binder1984}.



While theoretical descriptions do not provide a complete description, entropic simulations have proven to be a powerful tool to study phase transitions in several models \cite{Ferreira2021,Figueiredo2019,Fytas2011a,Tsai2009,Poulain2006}. Their findings have disclosed new results that need a scrutinized analysis to provide a coherent interpretation. An example is the ambiguous behavior of the phase transition for the spin-1 Baxter-Wu model\cite{Jorge2019}. A previous study showed that the critical behavior of quantities related to first and second-order phase transitions are equally satisfied. Furthermore, using a technique of tearing the densities of states curve, it was possible to observe that the ferromagnetic ground state undergoes a first-order phase transition and the ferrimagnetic ground state a second-order one.


Another feature evidenced by entropic simulations is the behavior of probability distributions of the microcanonical energy. For the Baxter-Wu model with a crystalline field \cite{Jorge2021}, this quantity presents an unusual noise classified as a finite-size effect. However, this statement has no proof up to now. Other models also showed similar behavior, such as the ZGB-like model\cite{Santos2022}, polymer chain \cite{Paul2008}, q Potts model \cite{Praveen2014}, $q=10$ and $20$ Potts\cite{Fiore2013a}, and the three-dimensional Ising model \cite{Ferrenberg1991}. To this extent, such behavior lacks a competent investigation.


The $q$-states Potts model has been extensively studied not only due to its theoretical interest but also for its practical applications. For $q\leq 4$, with no external magnetic field, the model can be exactly solved and is known to have a second-order phase transition with well-defined critical exponents that depends on $q$ \cite{Wu1982,Baxter1982}. For $q>4$, the phase transition is first order. Within the entropic simulations scenario, the Potts model is used to test improvements in the algorithm\cite{Landau2004,Cunha-Netto2008,Tsai2009,Vogel2013,Surungan2014,Caparica2015a,Caparica2015b}. Surungan \cite{Surungan2014} investigated the $q=4$ Potts model with the presence of invisible states using entropic simulations. They used the inverse of the microcanonical temperature, $\frac{\Delta S}{\Delta E}$, to determine the order of the phase transition. For the case where there is no invisible state in the lattice, the initial behavior of the curve does not match the expected result. The same is true for the spin 1 Baxter-Wu model \cite{Jorge2019}. 



Furthermore, when an intense and reverse external field acts to one of the $q$ states, the phase transition is analogous to the model with $q-1$ states with no external field applied \cite{Coutinho2006,Tsai2008}, i.e., the critical exponents and critical temperature are the same.


In this paper, we address the phenomena of phase transition of the $q=3$ Potts model using entropic simulations. More precisely, we will probe the logarithm of the density of states in the region close to the ground state energies by checking the configurations and analyzing the behavior of probability distribution of the microcanonical energy and the inverse microcanonical temperature. These quantities reveal the microcanonical behavior of the system and can clarify questions about the phase transition.

\section{Model}

The $q$-states Potts model has been at the forefront of scientific research in Statistical Mechanics since its inception.\cite{Potts1952}. It represents an extension of the ferromagnetic Ising model for the case $q>2$. In this model, the sites of a two-dimensional lattice can contain one of the $q$ possible states, represented by the spin variable $\sigma$, and interact ferromagnetically if two neighboring sites are equal. We can write the Hamiltonian as, 
\begin{equation}\label{eq:01}
\mathcal{H}=-J\sum_{\langle i,j\rangle} \delta_{\sigma_i\sigma_j}-H\sum_i \delta_{1,\sigma_i},
\end{equation}
where the first sum runs over the nearest neighbors and the second across the entire lattice. $\delta$ is the Kronecker delta, being 1 when the states of neighboring sites are equal and zero otherwise. $H$ represents the magnitude of an external field. It acts only on state 1, causing this state to be energetically favorable when $H>0$. Otherwise, it causes an energy increase due to the presence of this state in the lattice.

The ground state is formed by configurations where the lattice is completely filled with only one of the states. Its energy is independent of $q$ and corresponds to $E_g=-2J L^2$ for $H<0$. When $H>0$ the minimum energy is $E_g=-(2J + H) L^2$ and the lattice is filled by state 1. 

\section{Computational details}

To study the behavior of the density of states we use a modified version of the Wang-Landau algorithm \cite{Wang2001,Wang2001a}, called entropic simulation, due to its ability to estimate the logarithm of the density of states ($\ln g(E) )$) directly, which, except for a constant, is the Boltzmann entropy, $S(E) \equiv \ln g(E)$. This new version aims to improve the estimation of microcanonical averages and save computational time\cite{Caparica2012, Caparica2014,Ferreira2018}. The computational details are presented below. 

A new configuration is defined as changing the state of a site, chosen at random, to any of the possible states with equal probability. A Monte Carlo (MC) step is counted after $L^2$ new configuration attempts. The initial configuration is arbitrary, so we start from a ground state configuration, due to the ease of assigning the energy and order parameter to this configuration. Initially the density of states is not known, so we assign a unit to all possible energy levels of the system, $g(E)=1$, making $S(E)=0$. The functioning of the algorithm is based on the creation and monitoring of an energy histogram $H(E)$. The random walk through the energy space is from $E_{min}$ to $E_{max}$ with probability  
\begin{equation}
p(E \rightarrow E' ) = \min \{ \exp[(S(E) - S(E'))] , 1\},
\end{equation}
where $E'$ and $E$ are the energies of the new and current configurations, respectively. 
After a MC step $S(E)$ and $H(E)$ are updated, $S(E)\rightarrow S(E)+\ln f_i$ and $H(E)\rightarrow H(E) +1$, where $f_i$ is the so-called modification factor, which initial value is $f_0\equiv e= 2.71828 \cdots$. After a certain amount of MC steps the energy histogram flatness is checked. It is considered flat if $H(E)>0.8 \langle H \rangle $, for all energy levels, where $ \langle H \rangle$ is the mean value of the histogram. If it is not flat the random walk continues. Otherwise, we have an estimate for $S(E)$, completing a Wang-Landau level. To obtain $S(E)$ with greater precision the modification factor is refined $f_{i+1}=\sqrt{f_{i}}$ and the simulation continues with the random walk. From $f_i=f_7$ the microcanonical averages start to be accumulated. 

To finish a simulation run, we follow the evolution of the parameter 
\begin{equation}
  \varepsilon = |T_c(t)-T_c(0)|,
\end{equation}
where $T_c(0)$ and $T_c(t)$ are temperatures of the maximum of the specific heat obtained with $S(E)$ at the beginning of the WL level and when the flatness criterion is checked. If during an entire WL level this parameter remains less than $10^{-4}$ the simulation is halted. To save computational time, we start the verification at the eighth level, $f_7$. 
Identifying the Hamiltonian sums, Eq. \ref{eq:01}, as 
\begin{align}
E_1=&-\sum_{<i,j>} \delta_{\sigma_i\sigma_j} \\
E_2=&-\sum_i \delta_{1,\sigma_i},
\end{align} 
we can obtain the logarithm of the joint density of states $\ln g(E_1,E_2)$ \cite{Zhou2006,Kalyan2016,Ferreira2019,Jorge2021,Ferreira2021}, using the same simulation scheme, just exchanging $\ln g(E )$ by $\ln g(E_1,E_2)$, where $E=JE_1+HE_2$ . We can obtain $\ln g(E)_{J,H}$ from $\ln g(E_1,E_2)$ using the equation 
\begin{equation}
\ln g(E)_{J,H}=\ln \left( \sum_{E_1,E_2}\delta(E-JE_1-HE_2) e^{\ln g(E_1,E_2)} \right),
\end{equation}
where $\delta(E-JE_1-HE_2)$ is the Dirac delta that filters out the energies $E_1$ and $E_2$ that satisfy $E=JE_1+HE_2$. Thus, the estimation of the logarithm of the joint density of states allows us to investigate the microcanonical entropy for any external field. 

\section{Results}

\begin{figure}[b]
 \centering
 \includegraphics[scale=0.56, angle=-90]{ 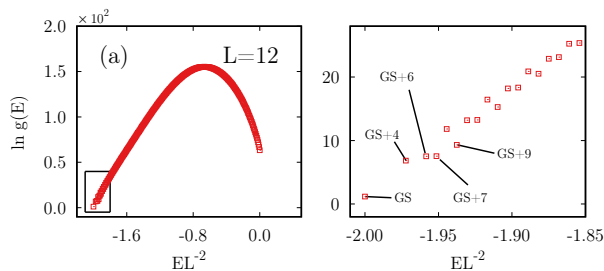}
 \caption{Inspection of the logarithm of the density of states for $L=12$. (a) Full density of states. (b) Detailed view of the logarithm of the density of states close to the ground state.  \label{fig01} }
\end{figure}
\begin{figure*}[t]
 \centering
 \includegraphics[scale=0.98, angle=-90]{ 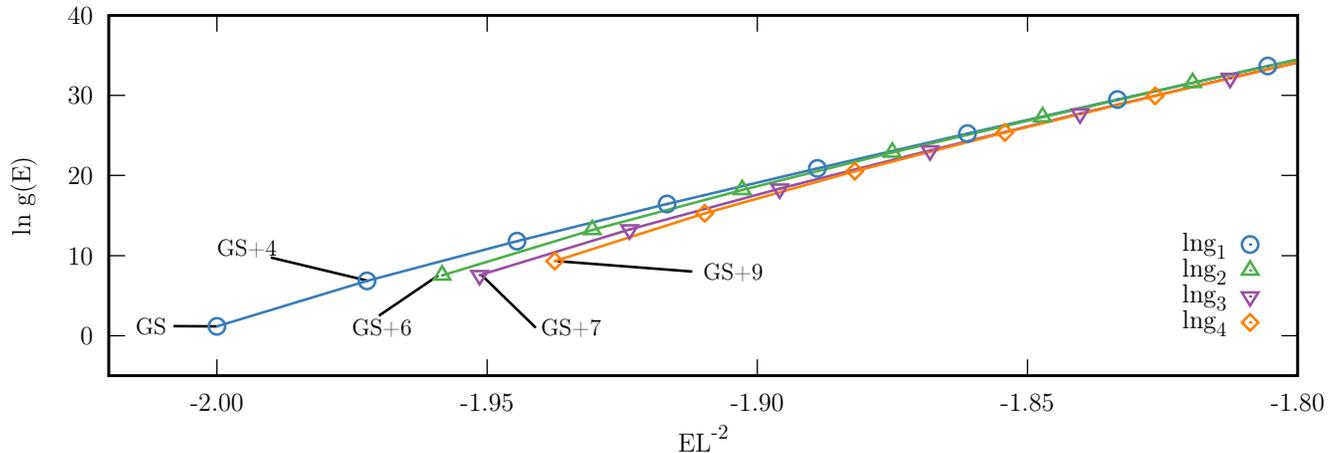}
 \caption{Logarithm of the density of states separated into four densities.  The labels $GS$, $GS+4$, $GS+6$, $GS+7$ and $GS+9$, $GS$ represents the ground state energy and the numbers are the changes in energy from the ground state. \label{fig03} }
\end{figure*}
\begin{figure}[t]
 \centering
 \includegraphics[scale=0.56, angle=-90]{ 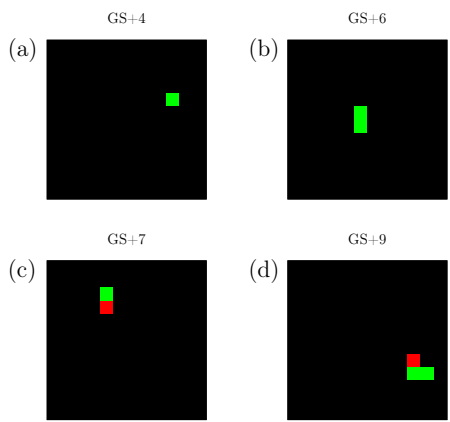}
 \caption{Configurations of the excited levels. \label{fig02} }
\end{figure}

\subsection{Case $H=0$}



Special attention must be paid to the logarithm of the density of states in the region close to the ground state. There one can notice a very peculiar behavior, which can be seen in Fig. \ref{fig01} for $L=12$. Looking at the logarithm of the density of states without much care one would say that there is nothing special, Fig. \ref{fig01}(a), but when we look closer, Fig. \ref{fig01}(b), we notice that there is an irregularity in the logarithm of the density of states.

This same behavior is found in the Ising model\cite{Beale1996} and spin 1/2 Baxter-Wu model. In order to understand what is going on, we investigated the configurations of the first four excited levels. They are labeled as $GS+4$, $GS+6$, $GS+7$ and $GS+9$. $GS$ represents the ground state energy and the numbers are the changes in energy from the ground state. The results for size $L=12$ can be seen in Fig. \ref{fig01}(b).

The first excited state has the configuration whose energy is equal to the ground state plus four ($GS+4$), as in Fig. \ref{fig02}(a). It is obtained by taking a ground state configuration and changing the state of one of the lattice sites. The same to the configuration whose energy is $GS+6$ (Fig. \ref{fig02}(b)): starting from the ground state configuration, two neighboring sites in the lattice are changed to the same state. If the neighboring sites are in different states, the energy becomes $GS+7$, as in \ref{fig02}(c). The energy configuration $GS+9$ is obtained by altering the states of three neighboring sites of the ground state configuration, with two equal states (Fig. \ref{fig02}(d)).

Due to the constraints on the positions of the chosen sites and states in the configurations $GS+6$, $GS+7$ and $GS+9$, the number of configurations with these energies will be lower than a configuration where they are not required. An example is the configuration where two non-neighboring sites are changed in the ground state configuration, generating a configuration whose energy is $GS+8$. This fact can be seen in the curve of the density of states in Fig. \ref{fig03}.  

When analyzing the following energies we observe the same behavior for these four configurations. For energy $GS+8$, the configuration keeps the exchanges distant, while for $GS+10$, $GS+11$ and $GS+13$ the chosen sites remain neighbors, changing only the states. We observe that configurations of similar characteristics are obtained by varying the energy of $\Delta E=4$ units starting from the first four excited states.

From these observations we can conclude that the cluster formations minimize the energy, i.e., when grouping equal states, the energy is less than keeping them apart. While the formation of clusters is avoided, the entropy is maximized, i.e., it increases the number of possible configurations. We see that different configurations are obtained by minimizing energy and maximizing entropy. By the energy-entropy argument, some phase transition must exist at some intermediate temperature\cite{Goldenfeld1992}. Such analysis suggests that other lattice models may also contain analogous behaviour of the density of states, such as the Ising model. 

We also realized that some of these energies can only be obtained if there is a specific dynamic to reach certain configurations. Apparently, each of these configurations develop independent densities of states that come together at a certain energy. In order to visualize this, we show the logarithm of the density of states starting from these four excited states and restricting the energy variation to 4. The result is shown in Fig \ref{fig03}, confirming our expectation. We will label $\ln g_1$, $\ln g_2$, $\ln g_3$ and $\ln g_4$ the logarithms of densities starting from the states $GS$, $GS+6$, $GS+7$ and $GS+9$, respectively.

We expect this behavior to be reflected in the probability distribution of energy
\begin{equation}
    P(E,T)=\frac{g(E)e^{-E/k_BT}}{Z},
\end{equation}
where $g(E)$ is the density of states, $k_B$ is the Boltzmann's constant and $Z$ is the partition function. Rewriting in terms of the logarithm of the density of states we have 
\begin{equation}
    P(E,T)=\frac{e^{-F/k_BT}}{Z},
\end{equation}
where $F=E-k_BT\ln g(E)$ is the Helmholtz free energy.

\begin{figure}[t]
 \centering
 \includegraphics[scale=0.56, angle=-90]{ 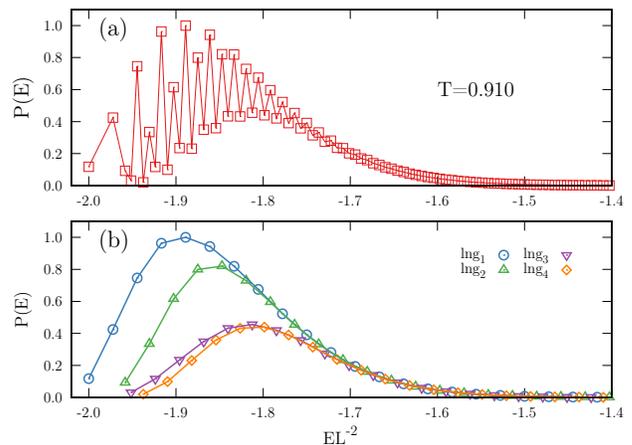}
 \caption{Distribution of energy probabilities without separating the logarithm of the density of states. \label{fig04} }
\end{figure}
To calculate the probability distribution of energy we use $k_B=1$ and neglect the partition function. Furthermore, we use a normalization factor $e^{-\lambda(T)}$, where $\lambda$ returns the maximum of $F$ in the fixed temperature energy interval, so that the probability distribution is written as  
\begin{equation}
    P(E,T)=e^{-F/k_BT-\lambda}.
\end{equation}
Due to the normalization factor, the maximum of the probability distribution is equal to unit. The value of the factor $\lambda$ is obtained using the logarithm of the density of states for all energy levels. The typical behavior of this quantity is shown in Fig. \ref{fig04}(a) for $T=0.91$. As can be seen, their behavior is peculiar when we use all energy levels.

\begin{figure}[t]
 \centering
 \includegraphics[scale=0.56, angle=-90]{ 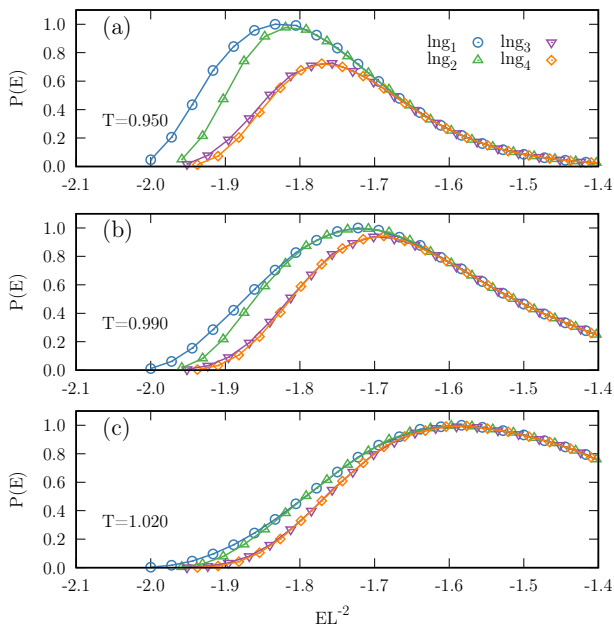}
 \caption{Probability distribution of energy separating the logarithm of the density of states into four curves shown for four different temperatures. In (a) $T=0.91$, (b) $T=0.95$, (c) $T=0.99$ and (d) $T=1.02$.\label{fig05} }
\end{figure}

Restricting the energy variation to $\Delta E=4$ and starting $P(E,T)$ from the energies of the first four excited states we obtain the curves shown in Fig. \ref{fig04}(b). In Fig. \ref{fig05} we show a sequence of temperatures. 

One can see that each curve of the logarithm of the density of states generates a well-behaved probability distribution, with a Gaussian like shape. The heights of the distributions show that, at low temperatures, the system is more likely to change its state to a configuration where the energy is not as low as possible. Regarding the configurations, the system does not present cluster formation. As the temperature increases, the presence of clusters is favored. Then the presence of clusters from several different states becomes more probable. As the temperature approaches the phase transition, the four curves tend to merge and their width tends to increase. The above description can be seen through the GIF that shows the most probable configurations as we increase the temperature (t-16.0.0.gif file, sent in attachment). To a better visualization of this procedure, check the supplemental material. 

\begin{figure}[t]
 \centering
 \includegraphics[scale=0.56, angle=-90]{ 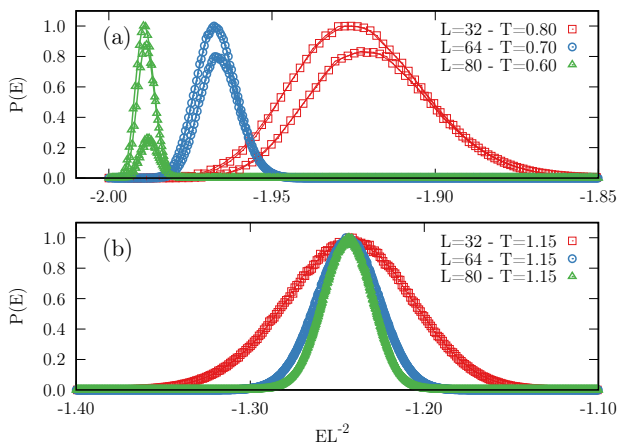}
 \caption{Dependency of the energy probability distribution with the lattice size $L=32,~64$ and $80$. In (a), the behavior for lattice sizes at different temperatures and in (b) at the same temperature.  \label{fig06} }
\end{figure}

To identify finite-size effects, we performed entropic simulations estimating the density of states for the sizes $L=~32,~64$ and $80$. In Fig. \ref{fig06} (a) we show the behavior of the energy probability distribution at temperatures $T=0.80,~0.70$ and $0.60$. In \ref{fig06}(b), showing the energy probability distributions for $T=1.15$ and $L=~32,~64$ and $80$, we realized that the larger the lattice size, the smaller the temperature at which the four energy probability curves collapse and that the collapsed curves, of all lattice sizes, are together after to the critical temperature of the system.  An explanation for this behavior follows from the fact that an energy variation $\Delta E=4$ becomes very small compared to the total energy range $2L^2$, yielding $\frac{\Delta E}{2L^2}=2L^{-2}$. At the thermodynamic limit we have $\lim_{L\rightarrow\infty} \frac{\Delta E}{2L^2} = 0$, so we expect the curves to collapse at very low energy, as the entire analysis is carried out in the region close to the ground state. However, as we are dealing with finite systems, all the effects observed here must be taken into account when tackling such quantities.

\begin{figure}[t]
 \centering
 \includegraphics[scale=0.56, angle=-90]{ 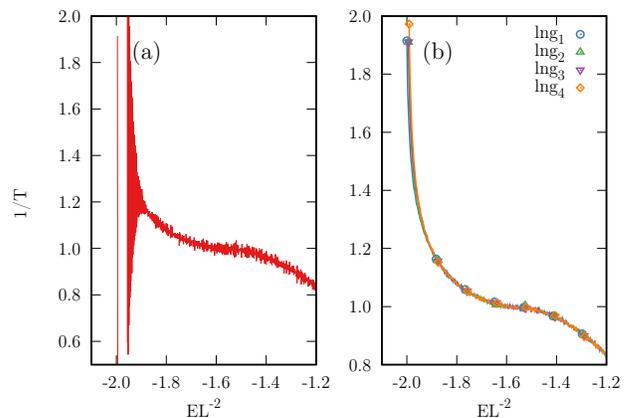}
 \caption{Inverse of the microcanonical temperature for $L=32$. (a) No separation of the logarithms of the densities of states and (b) separated.  \label{fig07} }
\end{figure}

Another quantity that also displays a significant difference under the above procedure is the microcanonical temperature $\frac{1}{T}=\frac{\Delta S}{\Delta E}$. In Fig \ref{fig07}(a) we show the behavior of this quantity without taking into account the different density of states curves. We noticed that at low energies the behavior of the curve does not agree with the expected one, $T\rightarrow 0\Rightarrow\frac{1}{T}\rightarrow \infty$. The same behavior was obtained by Surungan {\it et. al.}\cite{Surungan2014} for the $q=4$ Potts model and Fiore and DaSilva \cite{Fiore2013a} for the $q=10$ and $q=20$ Potts model. In fig \ref{fig07}(b), the curves represent the individual derivative of each of the four curves. In this case, the energy variation is $\Delta E=4$, regardless of the chosen density of states. Initially the curves are separated and, as in the case of the energy probability distribution, they merge at the transition temperature. The divergence $\frac{1}{T}\rightarrow \infty$ is only obtained in the thermodynamic limit\cite{Caparica2015b}, whose dependence is logarithmic.    


\subsection{Case $H<0$}
\begin{figure}[t]
 \centering
 \includegraphics[scale=0.56, angle=-90]{ 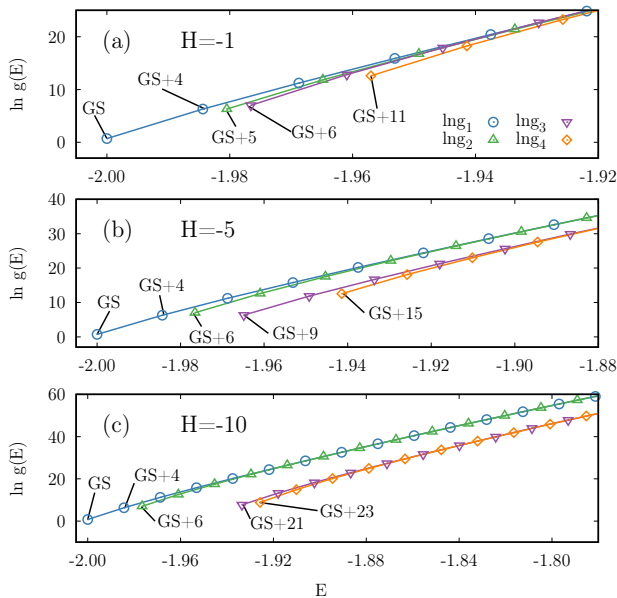}
 \caption{Logarithm of the density of states for q=3 and H=-1, -5 and -10 . \label{fig08} }
\end{figure}

Now we turn to what happens when we add a negative external field, i.e., keep out particles that interact with the field. Starting with $H=-1$, Fig. \ref{fig08}(a) shows the detail of the density of states with the four openings of the ground state. Here we have the levels $GS$, $GS+4$, $GS+5$, $GS+6$ and $GS+11$. The ground state is formed by one of the states that are not affected by the external field. The first excited state $GS+4$ is reached by changing the state from one site to one that does not interact with the field. If the new state interacts with the field it will correspond to the second excited state $GS+5$. 

The third excited state corresponds to the formation of clusters that do not interact with the external field. The beginning of the fourth density of states occurs when we have the conversion of three sites in the lattice. Two of them with states other than the ground state and not interacting with the field and one of them interacting with the field. Another possibility is the presence of three states interacting with the field and forming clusters.  

In Fig. \ref{fig08}(a), one can see that the line related to the ground state has a greater number of states. All the others, as they contains specifications regarding the shape and number of particles in each state, have a lower number of states. This behavior is valid up to the transition region between the phases of the system. It is noteworthy that the logarithm of the density of states in this interval is highly dependent on the number of sites, since the number of ways of rearranging the system with a fixed number of particles is proportional to the number of particles. The most likely configurations at different temperatures are shows in the supplementary material.




Investigating the logarithm of the density of states for $H=-5$ and $-10$ (see Fig. \ref{fig08} (b) and (c), respectively) we note that there is a well-defined separation between the lines. $GS+4$ and $GS+6$ are present in all fields, indicating that the energies do not depend on the field value, i.e., the states present in these configurations do not interact with the external field. On the other hand, $GS+9$ and $GS+15$ for $H=-5$ and $GS+21$ and $GS+23$ for $H=-10$ represent the entry of states interacting with the field, without and with formation of clusters, respectively. We note that the larger the field the further apart these lines become. 

\begin{figure}[t]
 \centering
 \includegraphics[scale=0.56, angle=-90]{ 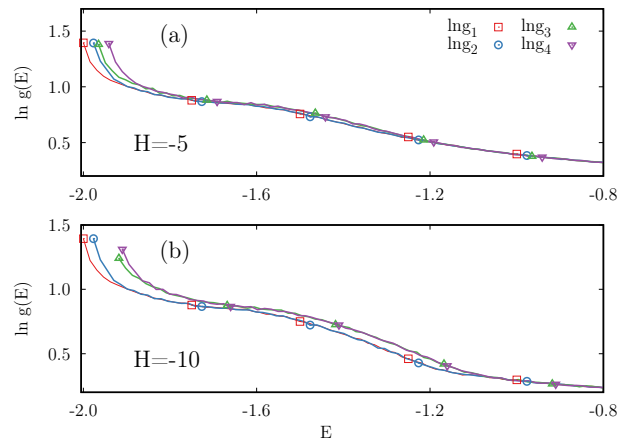}
 \caption{Inverse of microcanonical temperature for $q=3$, in (a) $H=-5$ and (b) $H=-10$.  \label{fig09}}
\end{figure}

In Fig. \ref{fig09} (a) and (b) we show the inverse of the microcanonical temperature for $H=-5$ and $-10$, respectively. One can see two distinct aspects. The first is the joining of the curves with states interacting with the field ($\ln g_3$ and $\ln g_4$), the joining of the non-interacting curves ($\ln g_1$ and $\ln g_2$), forming two distinct curves. When this happens, we have the formation of a plateau. The two curves seem to come together at the critical temperature, but come apart again. The second is the junction of the four curves. For $H=-10$, this effect is better visualized. As the curves meet, a second plateau appears. At this point, a lattice site is equally likely to contain any of the $q$ states, that is, the configuration is disordered. 

\subsection{Case $H>0$}

\begin{figure}[t]
 \centering
 \includegraphics[scale=0.56, angle=-90]{ 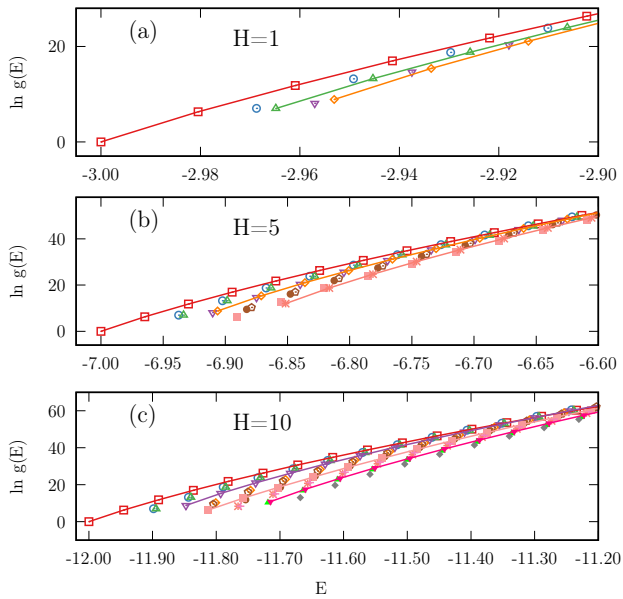}
 \caption{Logarithm of the density of states for q=3 and H=1, 5 and 10.\label{fig10} }
\end{figure}

When $H>0$, no finite-size effects is observed in the thermodynamic quantities. However, the logarithm of the density of states has a large number of independent lines, as can be seen in Fig. \ref{fig10}. When $H=1$ (see Fig \ref{fig10} (a)), we have a behavior similar to the previous cases: five independent curves can be obtained from the logarithm of the density of states. Now the variation between the points is 5 and the beginning of each curve are in $GS$, $GS+8$, $GS+9$, $GS+11$ and $GS+12$. Solely the lines corresponding to $GS$, $GS+9$ and $GS+12$ were presented for a better visualization, since the curves are very close to each other. A very useful comment about the configurations to this case are made in the supplementary material. 

For $H=5$, Fig \ref{fig10} (b), the number of curves increases to 9. The start of each curve happens at $GS$, $GS+16$, $GS+17$, $GS +23$, $GS+24$, $GS+28$, $GS+30$, $GS+31$ and $GS+38$ and the variation between the points is 9. For $H=10 $, Fig \ref{fig10} (c), we have 13 curves starting at $GS$, $GS+26$, $GS+27$, $GS+38$, $GS+48$, $GS+50$ , $GS+51$, $GS+60$, $GS+61$, $GS+62$, $GS+72$, $GS+73$ and $GS+85$ with variation of 14 between the points.

\begin{figure}[t]
 \centering
 \includegraphics[scale=0.56, angle=-90]{ 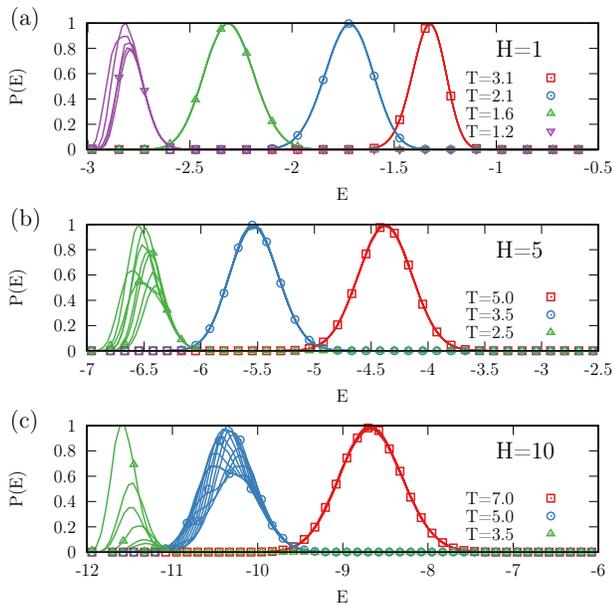}
 \caption{Probability distribution of energy separating the logarithm of the density of states into different curves shown for different temperatures. In (a) $H=1$, (b) $H=5$, and (c) $H=10$. \label{fig11} }
\end{figure}

In all cases analyzed for $H>0$ we observe that the curves tend to form a single curve, so that at the maximum temperature of the specific heat the most likely energy is the one where the curves of the logarithm of the density of states become equal. Observing the probability distribution of energy at different temperatures, shown in Fig. \ref{fig11}, we see that the different curves at low temperatures become equal at the maximum specific heat temperature.

\begin{figure}[t]
 \centering
 \includegraphics[scale=0.56, angle=-90]{ 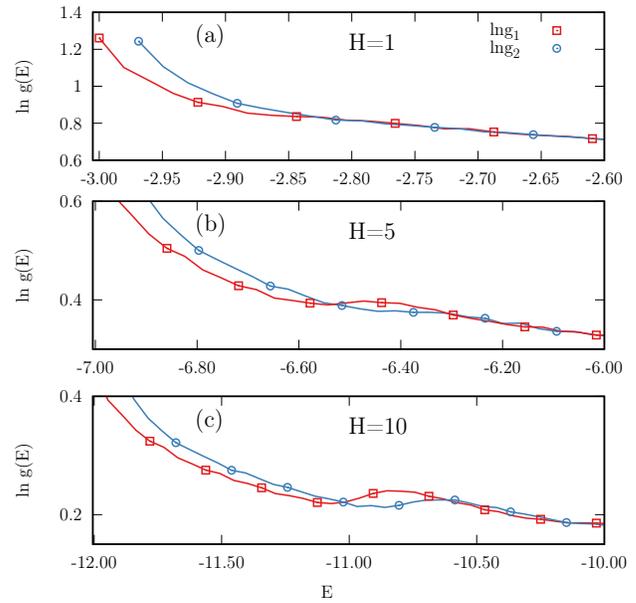}
 \caption{Configurations of the excited levels. \label{fig12} }
\end{figure}

The high number of curves interferes in the analysis of the behavior of the energy probability distribution, not allowing the visualization of double peaks in the curves with lower probabilities. To verify the occurrence of double peaks, we check the inverse of the microcanonical temperature for the first two curves of the logarithm of the density of states, shown in Fig. \ref{fig12}. In Fig. \ref{fig12} (a) for $H=1$, the temperature at which the curves meet is coherent with that of the maximum specific heat. When $H=5$, Fig. \ref{fig12} (b), the curve referring to $\ln g_1$ presents a first-order transition behavior while the second a second-order transition behavior, with the junction of the two curves taking place after the evidence of the two transitions. At $H=10$, the two curves show a first-order transition behavior. The omitted curves show a behavior similar to the ones presented in each case. The omission allows a better visualization of the first and second order behavior. 

\section {Conclusions}

In this work we have analyzed the logarithm of the density of states of the $q=3$ Potts model with an external field applied to one of the states. Using entropic simulations with two parameters, it was possible to observe the behavior of the microcanonical entropy, the probability distribution of the energy and the inverse of the microcanonical temperature for different external field values. The results show that the logarithm of the density of states near the ground state presents a behavior that indicates the presence of several curves that can be detached, which are related to the formation of clusters or not.

It was verified that the probability distribution and the inverse of the microcanonical temperature show the expected behavior when taking into account the different curves of the logarithm of the density of states. The microcanonical entropy is formed by different curves that come together after a certain temperature. We verified that, for $H=0$, this effect occurs for lattices of size $L=16,~32,~64$ and $80$, suggesting that it can be observed in all lattice sizes considered here.

For negative external fields, we observed the separation of two transitions: one related to the formation of clusters and other to the entry of states interacting with the external field. The first case is related to a critical transition and the second to a order-disorder one.

For positive external fields, where there no indication of a phase transition in the thermodynamic quantities, the microcanonical entropy possesses several independent curves, with the number of lines depending on the value of the external field. Thus, the probability distribution shows several curves that compose a gaussian-shaped envelope, and phase transition effects are not observed. The inverse of the microcanonical temperature showed that the curves independently present first-order phase transition signals at different temperatures.


Moreover, the visualization of this effect will help in the study of models with a more complex energy landscape, where the estimation of the microcanonical entropy is obtained in different energy ranges through simulation in parallel and then joined by observing the inverse of the microcanonical temperature. This procedure is called Replical Exchange Wang-Landau\cite{Vogel2014,Vogel2018,Boer2018}. As shown in this work, the energy region where the phase transition occurs is the same where the curves meet.

In summary, the logarithm of the density of states estimated by the entropic simulations is formed by different curves related to the formation of clusters and the phase transition characteristics may be present in situations where thermodynamic quantities such as specific heat and susceptibility do not present finite-size effects. 

\begin{acknowledgments}
We thanks the J. A. Plascak for comments. We thank the computer support from LaMCAD/UFG. L. S. Ferreira acknowledge the support by CNPq. 

\end{acknowledgments}

%
\end{document}


\title{An entropic approach to analyze phase transitions in the $q=3$ Potts model}

\author{L. S. Ferreira}
\email{ferreira\_s\_lucas@ufg.br}
\affiliation{Departamento de Física, Universidade Federal de Minas Gerais, Av. Antônio Carlos 6627, 31270-901, Belo Horizonte, Minas Gerais, Brazil}%

\author{L. N. Jorge}
\affiliation{Instituto Federal do Mato Grosso - Campus C{\'a}ceres, Av. dos Ramires s/n, 78200-000, C{\'a}ceres, MT, Brazil}

\author{C. J. DaSilva}
\affiliation{Instituto Federal de Goi\'as, Rua 76, Centro, Goi\^ania - GO, Brazil}

\author{A. A. Caparica}
\affiliation{Instituto de F\'{i}sica, Universidade Federal de Goi\'{a}s, Av. Esperan\c{c}a s/n, 74.690-900, Goi\^{a}nia, GO, Brazil}%

\date{\today}
             \maketitle
\section{Most likely configurations}

In this section we show how to obtain the most likely configurations using the procedure based on the Wang-Landau (WL) algorithm. In this type of simulation, the transition probability between the configurations is based on the inverse of the logarithm of the density of states, favoring a random walk through the energy space. Once the logarithm of the density of states is estimated, we can use it to promote an unrestricted walk through all possible energies. Furthermore, we can calculate the probability distribution of the temperature-dependent energy and energy parameters $J$ and $H$ as
\begin{equation}
    P(T,J,H)=g(E_1,E_2)e^{\frac{JE_1+HE_2}{k_BT}-\lambda},
\end{equation}
where $\lambda$ is the normalization factor used in equation 9 in the body of the article.

Combining these two characteristics, one can obtain the configurations associated with the most probable energies. By fixing $T$, $J$, and $H$ one can find the energy that maximizes the probability distribution, $E_1^{max}$ and $E_2^{max}$, and then use the WL acceptance probability for the configurations of these energies. A more detailed explanation of the algorithm can be found in Ref \cite{Ferreira2021}.

\section{Animated figure}

This procedure was performed at different temperatures from $T=0.1$ to $T=2.0$ and $5$ configurations were obtained for each temperature. Therefore, the sequence of configurations were used to create an animated figure (GIF). The sequence of images shown are just a representation of the most likely configurations and should not be confused with those obtained by the Metropolis algorithm, which represent the evolution of the system.

\subsection{$H=0$}

The \verb|t-16.0.0.gif| file, sent in the attachment, shows the most likely configurations for the case $H=0$. The states $q=1,~2$ and $3$ are represented by the colors black, red, and green, respectively. We observed that the evolution of the system with the increase of temperature, starting from the ground state, occurs with the passage from configurations without the formation of agglomerates to configurations with agglomerates.
\begin{figure}[!ht]
    \centering
    \includegraphics[scale=0.56,angle=-90]{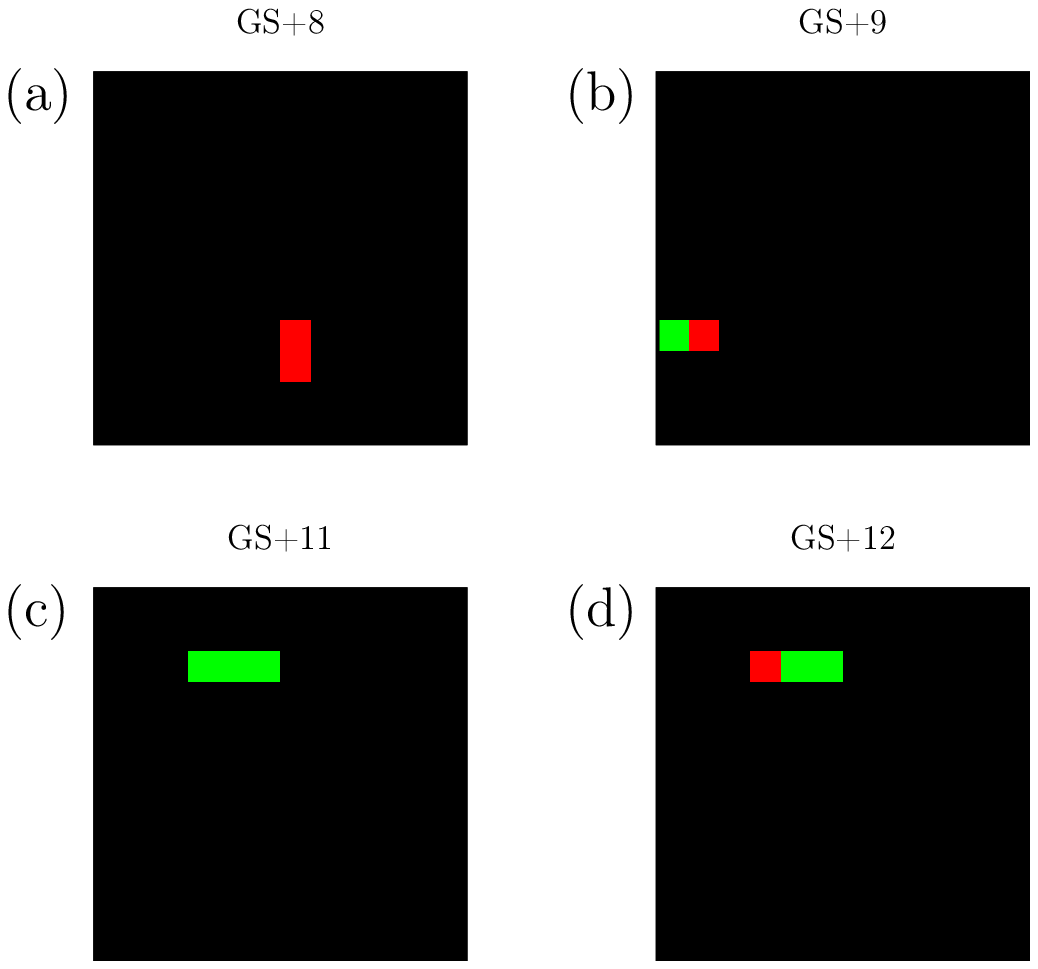}
    \caption{Configurations related to the beginning of the last four lines in the logarithm of the density of states for $H=1$ com $L=12$.}
    \label{fig03s}
\end{figure}

\begin{figure}[!ht]
    \centering
    \includegraphics[scale=0.56,angle=-90]{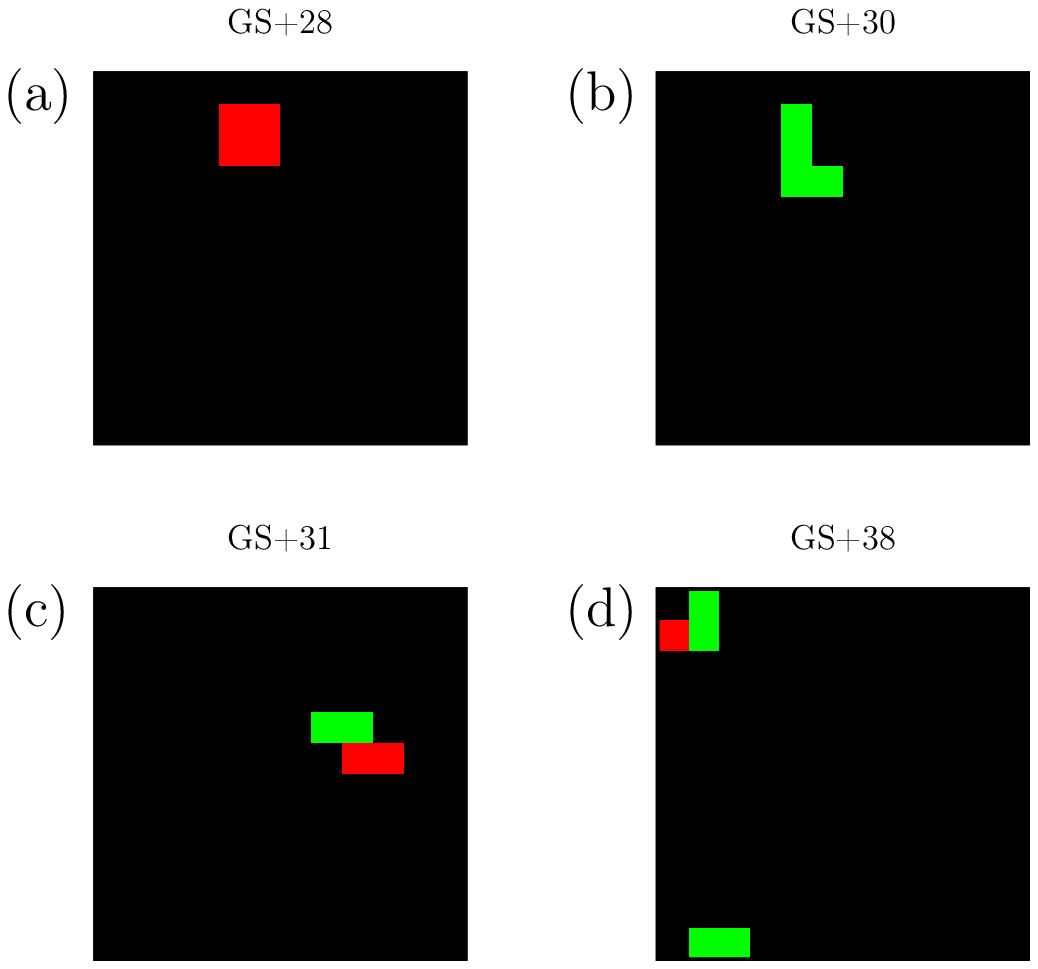}
    \caption{Configurations related to the beginning of the last four lines in the logarithm of the density of states for $H=5$ with $L=12$.}
    \label{fig04s}
\end{figure}

\subsection{Case $H<0$}

GIF \verb|t-16.-1.0.gif| shows the most likely configurations at different temperatures. We noticed that, up to the temperature $T=2.0$, the system still has blocks of states that do not interact with the external field. At this temperature, a disorder configuration was expected, since the transition occurred in $T\sim 1$. We can conclude that the transition to disorder has not yet occurred. A sign of the occurrence of this transition should appear in one of the thermodynamic quantities, such as specific heat or susceptibility.

\begin{figure}[!ht]
    \centering
    \includegraphics[scale=0.56,angle=-90]{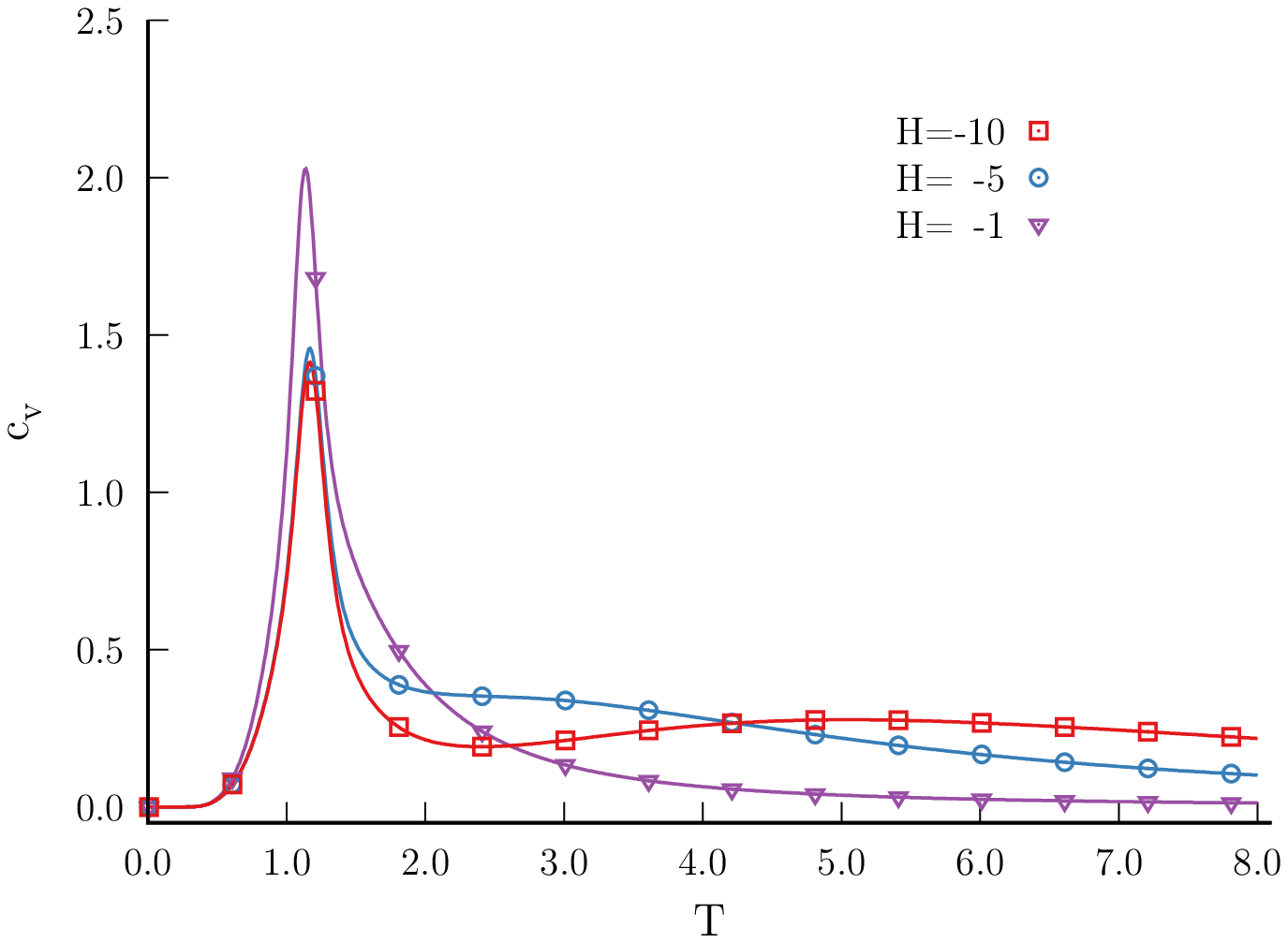}
    \caption{Specific heat behavior for $H=-1,~-5$ and $-10$ with $L=12$.}
    \label{fig01s}
\end{figure}

As shown in Fig. \ref{fig01s}, the specific heat for $H=-1$ has no visible effect of a second phase transition. However, for larger fields we notice a second maximum. 

In GIFs \verb|t-16.-5.0.gif| and \verb|t-16.-10.0.gif|, we clearly see the beginning of the formation of clusters close to the pseudo-critical temperature. This pattern is maintained until the entry of non-interacting with the field states, whose apex occurs near the temperature of the second maximum of the specific heat.
\subsection{Case $H>0$}

\begin{figure}[!ht]
    \centering
    \includegraphics[scale=0.56,angle=-90]{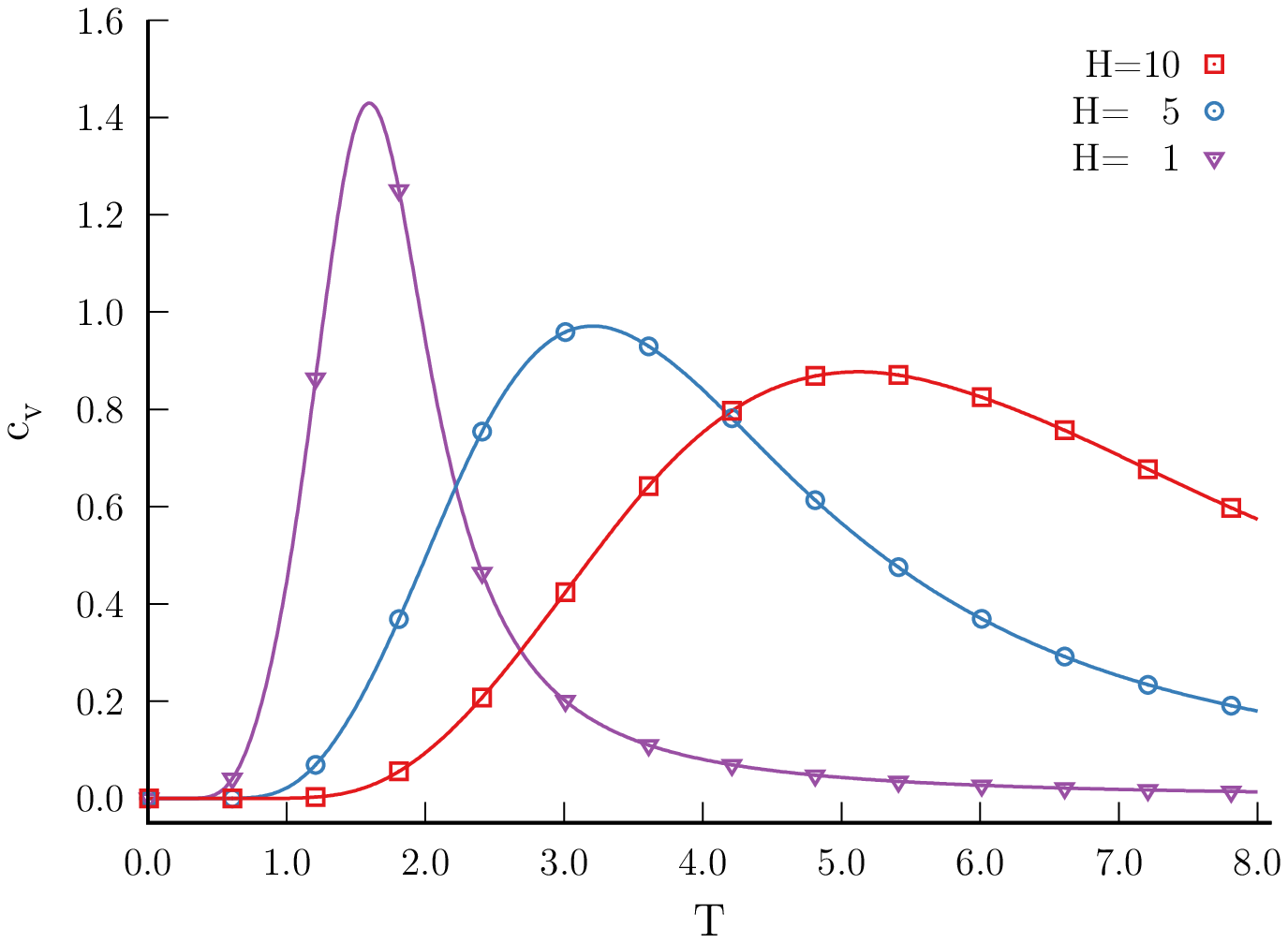}
    \caption{Specific heat behavior for $H=1.~5$ and $10$ with $L=12$.}
    \label{fig02s}
\end{figure}

\begin{figure}[!t]
    \centering
    \includegraphics[scale=0.56,angle=-90]{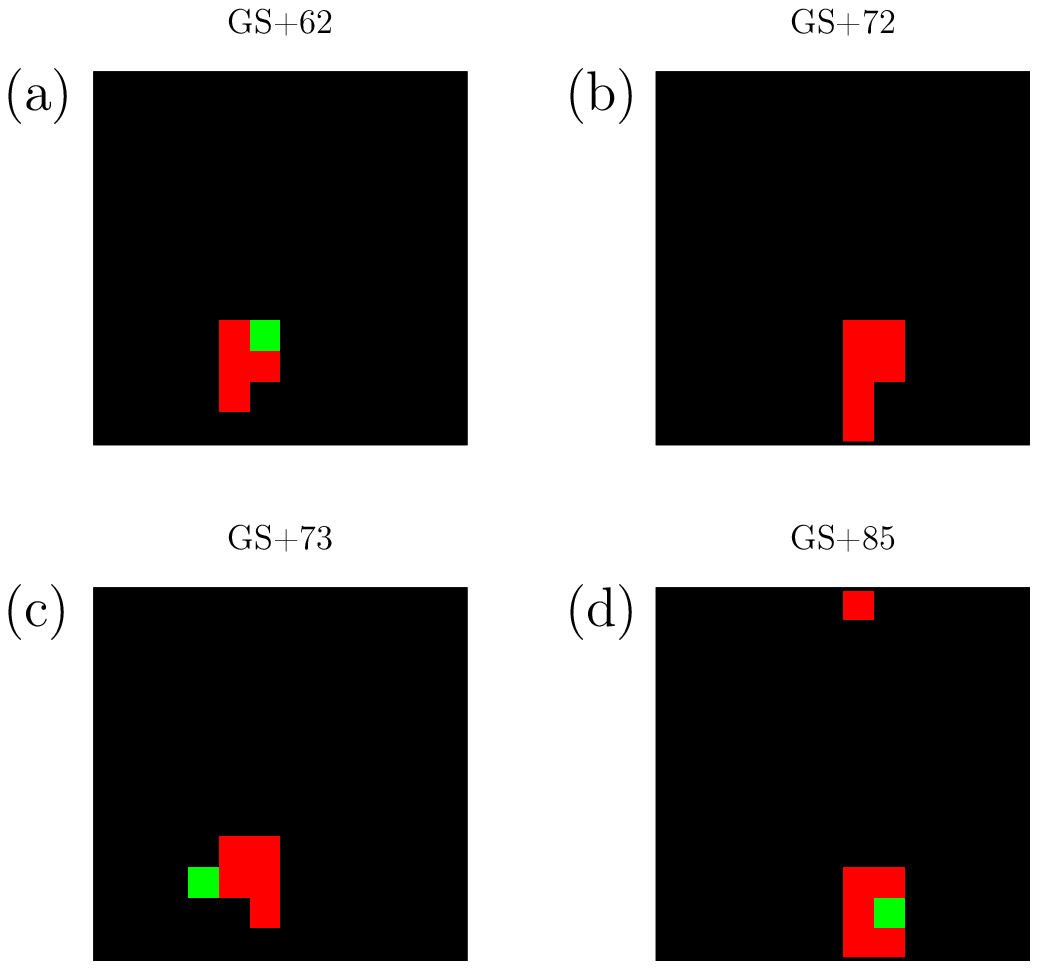}
    \caption{Configurations related to the beginning of the last four lines in the logarithm of the density of states for $H=10$ with $L=12$.}
    \label{fig05s}
\end{figure}

When the field is positive, the division of the density of states is quite pronounced and increases with the field. The maximum of the specific heat is rounded off (see Fig\ref{fig02s}). Although there is no finite size effect, there is a temperature shift from the specific heat maximum with increasing field. We observed that the patterns of the configurations are repeated, that is, when we increase the field the new lines that appear are due to different cluster patterns. The lines related to the patterns observed in configurations with lower field are maintained. Choosing $H=1.~5$ and $10$, we only show the configurations related to the beginning of the last four lines of the logarithm of the density of states for each field. Doing that, we have access to all patterns related to the beginning of lines up to the $H=10$ field. The beginning of all lines shows that they are related to a type of cluster formation as can be seen in Figs. \ref{fig03s}, \ref{fig04s} and \ref{fig05s}.

Thus, if we set the external field to a positive value, the increase in temperature will promote transitions, firstly, to configurations where only one site is changed, then configurations with cluster formation gradually emerge with temperature. For smaller fields, the cluster patterns are more mixed, as the energy fluctuation must be greater in these fields, which matches the behavior shown in Fig.\ref{fig02s}. For larger fields, the transition between different cluster states takes place at different temperatures. Thus, we assume that the specific heat curve with a rounded maximum is due to different transitions between the clusters. Its intensity is lower due to the reduced amount of probable configurations at the same temperature. The analysis described here is in accordance with the most likely configurations represented in GIFs \verb|t-16.1.0.gif| and \verb|t-16.10.0.gif|. 

\bibliography{referencias.bib}